\documentclass[fleqn,usenatbib]{rasti}

\usepackage{newtxtext,newtxmath}
\usepackage[T1]{fontenc}

\DeclareRobustCommand{\VAN}[3]{#2}
\let\VANthebibliography\thebibliography
\def\thebibliography{\DeclareRobustCommand{\VAN}[3]{##3}\VANthebibliography}

\usepackage{graphicx}	
\usepackage{amsmath}	

\def\msun{M$_{\odot}$}

\def\shark{\textsc{Shark}}

\title[Stellar Masses from Photometry via Neural Network]{Fast and Accurate Stellar Mass Predictions from Broad-Band Magnitudes with a Simple Neural Network: Application to Simulated Star-Forming Galaxies}

\author[E. Elson]{
E. Elson$^{1}$\thanks{E-mail: eelson@uwc.ac.za (EE)}
\\
$^{1}$Department of Physics \& Astronomy, University of the Western Cape, Robert Sobukwe Road, Bellville 7535, South Africa\\
}

\pubyear{\the\year{}}

\begin{document}
\label{firstpage}
\pagerange{\pageref{firstpage}--\pageref{lastpage}}
\maketitle

\begin{abstract}
A simple, fully connected neural network with a single hidden layer is used to estimate stellar masses for star-forming galaxies. The model is trained on broad-band photometry - from far-ultraviolet to mid-infrared wavelengths - generated by the Semi-Analytic Model of galaxy formation (\shark), along with derived colour indices. It accurately reproduces the known \shark\ stellar masses with respective root-mean-square and median errors of only 0.085 and $\sim0.1$~dex over the range $10^8$–$10^{11}$~\msun. Analysis of the trained network’s parameters reveals several colour indices to be particularly effective predictors of stellar mass. In particular, the $FUV - NUV$ colour emerges as a strong determinant, suggesting that the network has implicitly learned to account for attenuation effects in the ultraviolet bands, thereby increasing the diagnostic power of this index. Traditional methods such as spectral energy distribution fitting, though widely used, are often complex, computationally expensive, and sensitive to model assumptions and parameter degeneracies. In contrast, the neural network relies solely on easily obtained observables, enabling rapid and accurate stellar mass predictions at minimal computational cost. \textcolor{black}{The model derives its predictions exclusively from patterns learned in the data, without any built-in physical assumptions (such as stellar initial mass function)}. These results demonstrate the utility of this study's machine learning approach in astrophysical parameter estimation and highlight its potential to complement conventional techniques in upcoming large galaxy surveys.
\end{abstract}

\begin{keywords}
galaxies: evolution -- methods: numerical
\end{keywords}


\section{Introduction}
Understanding the stellar mass of galaxies is paramount to unravelling the processes that drive galaxy formation and evolution. Stellar mass not only acts as a key parameter in scaling relations—such as the stellar–halo mass relation \citep{behroozi_2013, moster_2013} and the star formation main sequence \citep{noeske_2007, speagle_2014}—but also informs us about the cumulative history of star formation and feedback within galaxies.

Traditionally, stellar masses of galaxies are estimated via spectral energy distribution (SED) fitting, a technique that models the integrated light of a galaxy over a broad wavelength range using synthetic stellar population templates. Although widely applied, SED fitting is fundamentally limited by inherent degeneracies among key parameters such as stellar age, metallicity, dust attenuation, and star formation history \citep{pforr_2012, conroy_2013}.  For example, the effects of an older, dust-obscured population can mimic those of a younger, less reddened one, leading to uncertainties of approximately 0.2–0.3 dex in the derived stellar masses. \textcolor{black}{These ambiguities are compounded by additional factors such as the choice of initial mass function, stellar population synthesis model, dust attenuation law, and the parameterisation of star formation histories, all of which can introduce significant systematic biases into the derived stellar masses \citep{mobasher_2015, leja_2019}}. Furthermore, the interplay between data quality, model accuracy, and fitting techniques used to derive physical galaxy properties can be complex (e.g.,~\citealt{walcher_2011, courteau_2014}). Consequently, even when high-quality multi-wavelength data are available, the derived stellar masses remain susceptible to these uncertainties, complicating efforts to trace galaxy evolution accurately.  \textcolor{black}{Numerous theoretical investigations have systematically explored how variations in modelling assumptions—including those governing star formation history, dust geometry, and IMF—affect inferred stellar masses and introduce both random and systematic uncertainties \citep{pforr_2012, pacifici_2015, leja_2019}.}

Recent advances in machine learning have opened alternative avenues for estimating galaxy properties, offering powerful data-driven techniques that complement or surpass traditional model-based approaches. Unlike SED fitting, which relies on pre-defined stellar population templates and assumptions about galaxy evolution, machine learning methods learn patterns directly from data, enabling fast and flexible predictions. Techniques such as artificial neural networks, random forests, and convolutional neural networks have been successfully applied to infer key galaxy parameters—including stellar mass, star formation rate, and photometric redshift—from multi-band photometry and spectroscopy (e.g.~\citealt{disanto_2018, pasquet_2019}). These methods reduce computational costs and mitigate biases arising from modelling assumptions, making them well-suited to the analysis of large galaxy surveys.

Previous studies aimed at predicting galaxy stellar masses using machine learning have often employed complex architectures—such as deep convolutional neural networks (CNNs)—that ingest high-dimensional inputs like 2D images, pixel-level spectral energy distributions, or spatially resolved maps (e.g.~\citealt{kamdar_2016, lovell_2022}).  \textcolor{black}{CNNs are particularly well suited to image data, where spatial correlations between pixels carry important information, such as morphological features or structural priors that can aid in tasks like photometric redshift estimation or morphological classification.}  While powerful, these approaches are computationally intensive and require detailed data that may not be readily available for large galaxy samples. In contrast, this study demonstrates that a very simple fully connected artificial neural network (ANN), consisting of just a single hidden layer, can produce accurate stellar mass estimates using only broad-band photometric measurements \textcolor{black}{and redshifts}. These inputs are observationally inexpensive to obtain and represent standard data products from forthcoming wide-field surveys such as LSST\footnote{Large Synoptic Survey Telescope \citep{ivezic_2019}.}, highlighting the practicality and efficiency of this approach for future large-scale galaxy studies.

The structure of this paper is organised as follows. Section~\ref{section_data} provides a detailed overview of the primary datasets used to train and test the ANN. The architecture of the ANN, along with the methods employed for training its parameters, is discussed in Section~\ref{section_ANN}. Sections~\ref{section_results} and \ref{section_discussion} present the results from the ANN's outputs together with a brief discussion, while Section~\ref{section_dynamics} delves into the dynamics of the network. Finally, Section~\ref{section_conclusions} provides a concise conclusion of this study.

\section{Data}\label{section_data}

The primary data set for this study is derived from the semi-analytic model of galaxy formation, \shark\ \citep{lagos_2018}.  \textcolor{black}{In line with many semi-analytic models,} \shark\ implements two principal approaches for gas cooling: one that ties the cooling timescale to the halo’s dynamical time \citep[following][]{croton_2006}, and another that determines cooling based on the time available for gas to lose heat \citep{benson_2010}. For regulating gas accretion via active galactic nuclei (AGN), \shark\ incorporates models ranging from a simple threshold-based feedback prescription to a more detailed approach linking black hole growth with energy output \citep{bower_2006, croton_2016}. Additionally, \shark\ models feedback from stars and photo-ionisation, accounting for both supernova-driven outflows \citep{lagos_2013, muratov_2015} and the suppression of star formation in low-mass haloes due to early ionising radiation \citep{lacey_2016, sobacchi_2013}. Star formation is implemented through multiple prescriptions that either use interstellar pressure to estimate molecular gas fractions \citep{blitz_2006}, or rely on simulation-informed and theoretical treatments \citep{gnedin_2014, krumholz_2009, krumholz_2013} of the atomic-to-molecular phase transition. These various implementations allow \shark\ to accurately reproduce key observational constraints, including the stellar mass function, stellar–halo mass relation at $z=0$–4, the evolution of the star formation rate density, and local gas scaling relations.

\citet{lagos_2019} extend \shark\ by coupling it with the \textsc{Prospect} SED generation tool \citep{robotham_2020} to model the multi-wavelength emission of galaxies from the far-ultraviolet (FUV) to the far-infrared (FIR). Galaxy attenuation is modelled using an empirical relation to estimate dust mass from gas metallicity and mass, with attenuation curves drawn from EAGLE simulations and intrinsic properties of \shark\ galaxies. This approach successfully reproduces a range of observational benchmarks, including various luminosity functions, FUV–FIR galaxy number counts, and the integrated cosmic SED out to $z=1$.

To account for both the attenuation of stellar light and its re-emission in the mid-to-far infrared, \citet{lagos_2019} adopt the \citet{dale_2014} dust templates, assuming negligible AGN contributions. While \textsc{Prospect} allows for AGN emission to be included when fitting galaxy SEDs, this component is not implemented in \shark\ due to the additional complexity required to scale AGN SED templates with physically meaningful AGN properties.  \textcolor{black}{It should be noted, however, that other semi-analytic models do include prescriptions for AGN emission at various wavelengths, typically by linking black hole accretion rates to empirical or theoretical AGN templates \citep[e.g.][]{bower_2006, fanidakis_2012}.}  The present study, therefore, focuses on star-forming galaxies whose infrared emission is not expected to be significantly affected by AGN activity. To select such galaxies from \shark, constraints are placed on their bulge-to-total mass ratio ($B/T$). \citet{morselli_2017} establish that galaxies on the star formation main sequence typically exhibit low $B/T$ values (around 0.15), while those deviating from the main sequence (yet still forming part of it) can reach values of approximately 0.65. Accordingly, the \shark\ sample is restricted to galaxies within this range. Furthermore, only galaxies with redshift $z<0.3$ are considered, consistent with selections such as those made in the GALEX–SDSS–WISE Legacy Catalogue \citep{salim_2016}.

\begin{figure}  
	\includegraphics[width=1\columnwidth]{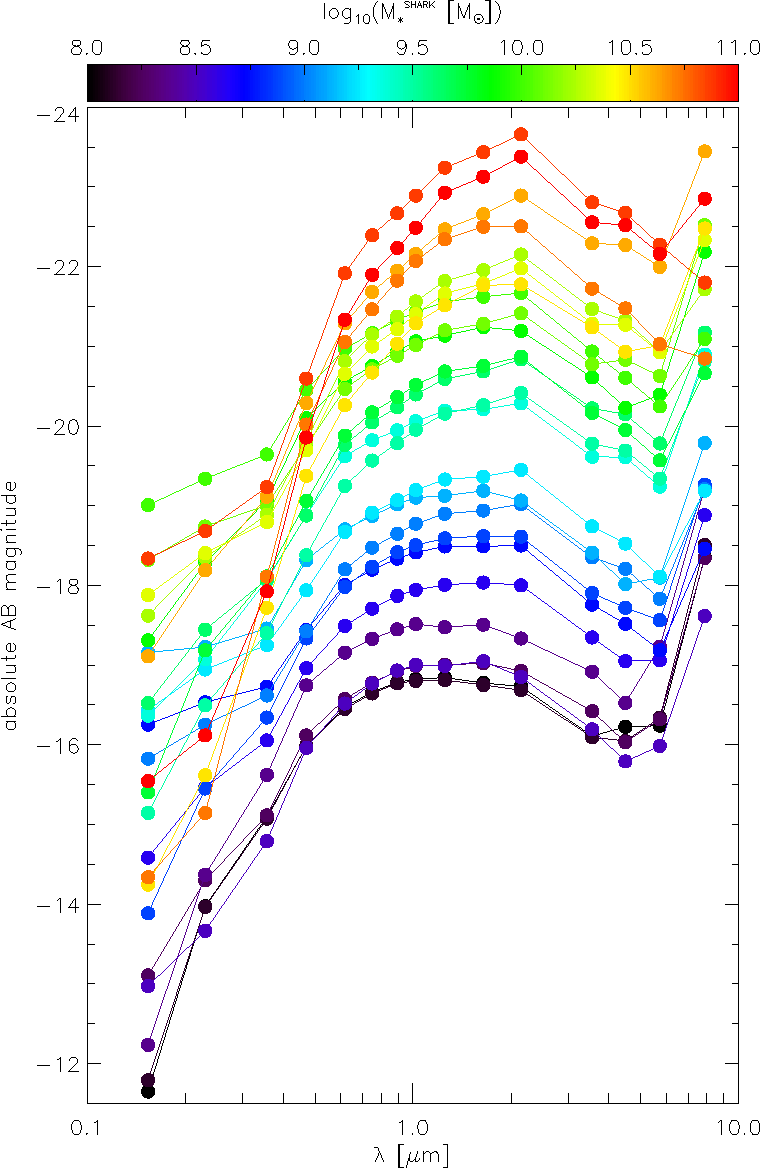}  
   \caption{Examples of \shark\ galaxy spectra. The value of \shark-evaluated $\log_{10}(M_*/M_{\odot})$ for each galaxy is indicated in the legend. These spectra, along with a set of associated colour indices, form the input features for the ANN, which is trained to predict the stellar mass of a galaxy based on these inputs. }  
    \label{spectra}  
\end{figure}

For each selected \shark\ galaxy, apparent AB magnitudes are compiled in the following filters: GALEX $FUV$ and $NUV$; SDSS $u$, $g$, $r$, $i$, $z$; VISTA\footnote{Visible and Infrared Survey Telescope for Astronomy \citep{vista}.} $Y$, $J$, $H$, $Ks$; and Spitzer $W1$, $W2$, $W3$, $W4$. These magnitudes are converted to absolute magnitudes using each galaxy’s observed redshift (incorporating both cosmological and peculiar components). Figure~\ref{spectra} shows examples of spectra of galaxies spanning the full stellar mass range of the \shark\ sample used in this study. In addition to broad-band magnitudes, a well-suited set of colour indices is also used for stellar mass estimation: $u-g$, $u-r$, $g-r$, $r-i$, $i-z$, $r-Ks$, $g-Ks$, $i-Ks$, $J-H$, $H-Ks$, $W1-W2$, $W2-W3$, $FUV-NUV$, $NUV-u$.

\begin{figure*}  
	\includegraphics[width=2\columnwidth]{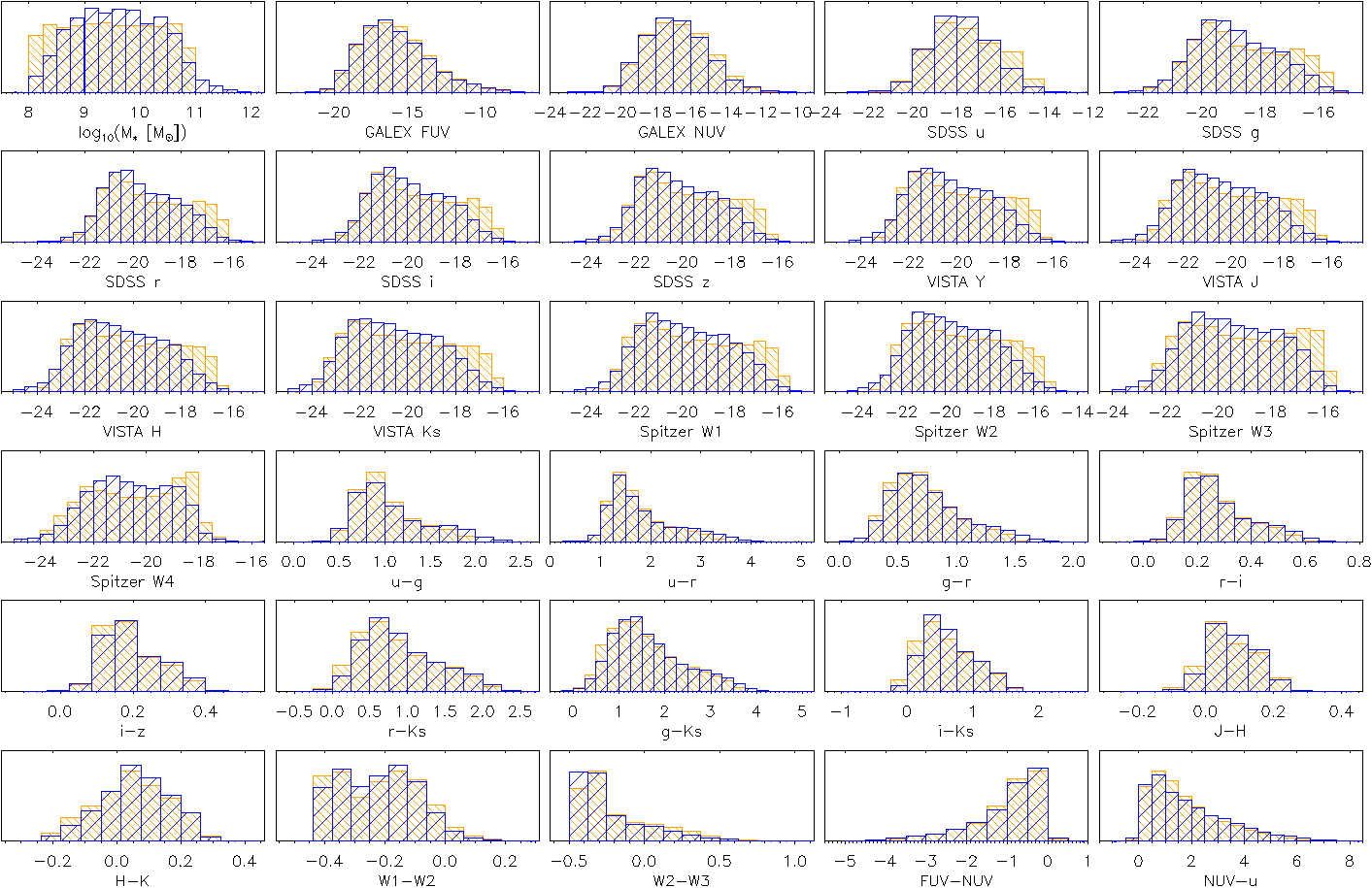}  
    \caption{Distributions of galaxy properties used to train the ANN. All panels except the first show data in units of AB magnitudes. Orange histograms represent the training set, constructed by selecting galaxies with $0.15 \leq B/T \leq 0.65$ to isolate systems near the star formation main sequence, followed by re-sampling to ensure uniform distribution in $\log_{10}M_*^\mathrm{SHARK}$. For comparison, blue histograms show the full \shark\ sample without any $B/T$ selection. While this cut excludes the most massive galaxies—whose infrared emission is likely affected by AGN activity—the overall distributions of magnitudes and colours remain largely unchanged. Both samples include only those galaxies with redshifts $z < 0.3$. Before being input into the ANN, all features are normalised to the range [0, 1].}  
    \label{distributions}  
\end{figure*}

To reduce the influence of extreme values, the distributions of all selected input quantities are trimmed to remove long tails. Additionally, galaxies are resampled to ensure approximately uniform distribution in logarithmic \shark-evaluated stellar mass ($\log_{10}M_*^\mathrm{SHARK}$), thereby preventing the ANN from being biased towards more numerous galaxies in particular mass ranges. From this curated data set, subsets of 8213 and 3520 galaxies are drawn for use as the training and test sets, respectively. Figure~\ref{distributions} shows the resulting property distributions for the training set—including $\log_{10}M_*^\mathrm{SHARK}$—alongside those of a representative subset of all \shark\ galaxies with $z < 0.3$, before the application of $B/T$ cuts and stellar mass uniformisation. After independently normalising all features to the range [0, 1], the final set of 29 input features is fed into the ANN used in this study to predict stellar masses.

\section{Artificial neural network}\label{section_ANN}
The artificial neural network (ANN) developed in this study is intentionally simple, demonstrating that accurate results can be achieved without resorting to deep, multilayered architectures. The network comprises an input layer with 29 neurons (indexed by \( i \)), each representing a magnitude or colour feature of a galaxy; a single hidden layer with 15 sigmoid neurons (indexed by \( j \)); and an output layer consisting of a single rectified linear unit (ReLU) neuron. It is a feed-forward network, with data flowing unidirectionally from the input to the output layer through fully connected neurons. A graphical representation of the network is shown in Fig.~\ref{ANN_fig}.

For the \( j \)-th hidden-layer neuron with bias parameter \( b_j^\mathrm{HL} \), the weight \( w_{ji}^\mathrm{HL} \) connects it to the \( i \)-th input-layer neuron. These parameters are used by each hidden-layer neuron to linearly combine the signals from all 29 input-layer neurons to form  
\begin{equation}
z_j^\mathrm{HL} = \sum_i w_{ji}^\mathrm{HL} x_i + b_j^\mathrm{HL}.
\end{equation}  

Each hidden-layer neuron then applies the sigmoid activation function to generate an output  
\begin{equation}
\sigma_j^\mathrm{HL} = \frac{1}{1 + \exp\left(-z_j^\mathrm{HL}\right)}.
\end{equation}  
This defines a sigmoid neuron \citep{rumelhart_1986, bishop_1995}, which maps the input to a nonlinear output ranging between 0 and 1.  

The network's output neuron is chosen to be of the ReLU type \citep{glorot_2011} to ensure that its output remains a continuous linear variable. Its bias parameter is $b^\mathrm{out}$, while $w_j^\mathrm{out}$ represents the weight connecting it to the \( j \)-th hidden-layer neuron. The neuron's inputs are linearly combined to form
\begin{equation} 
z^\mathrm{out} = \sum_j w_j^\mathrm{out} \sigma_j^\mathrm{HL} + b^\mathrm{out}, 
\end{equation}
\textcolor{black}{which represents the base-10 logarithm of a galaxy's stellar mass.}

Since the ReLU function preserves linearity for positive values, this combination serves as both the neuron's linear output—ranging between 0 and 1—and the overall output of the network.

The bias and weight parameters of the network are trained using a stochastic gradient descent algorithm \citep{robbins_1951, kiefer_1952}. At each epoch, a randomly selected subset of 50 galaxies from the training set is used to update the parameters. Given a parameter $p$\footnote{That is, one of $b_j^\mathrm{HL}$, $w_{ji}^\mathrm{HL}$, $b^\mathrm{out}$ or $w_j^\mathrm{out}$.}, its value is adjusted according to
\begin{equation} 
p \rightarrow p - \eta \frac{\partial z^\mathrm{out}}{\partial p}, 
\end{equation}
where $\eta=0.25$ is the learning rate.  The objective is to minimise the quadratic cost function
\begin{equation} 
Q(b_j^\mathrm{HL}, w_{ji}^\mathrm{HL}, b^\mathrm{out}, w_j^\mathrm{out}) = 0.5 (z^\mathrm{out} - \log_{10}M_*^\mathrm{SHARK})^2,
\end{equation}
where $M_*^\mathrm{SHARK}$ represents the known (i.e., \shark-evaluated) stellar mass of the input galaxy.  

\textcolor{black}{Some experimentation was conducted with the hyperparameters of the network, including the number of hidden-layer neurons, the learning rate, and the form of the cost function. Networks with 10 to 30 hidden-layer neurons were tested, and although larger networks converged slightly faster, the prediction accuracy remained largely unchanged. Using only 15 neurons was found to be sufficient, and was preferred in line with the study’s aim of demonstrating the effectiveness of a simple architecture. Similarly, a range of learning rates between 0.1 and 0.5 was trialled, with $\eta = 0.25$ yielding sufficiently stable and efficient convergence. Finally, although a cross-entropy cost function was also tested, it did not outperform the simple quadratic form, which was ultimately adopted for its direct interpretation in the context of regression.}

At the start of training, all bias parameters are drawn from a Gaussian distribution with a mean of 0 and a standard deviation of 0.1. Similarly, the weight parameters $w_{ji}^\mathrm{HL}$ and $w_j^\mathrm{out}$ are drawn from Gaussian distributions with standard deviations of $1/\sqrt{29}$ and $1/\sqrt{15}$, respectively. This initialisation helps maintain stable gradient propagation during training and, hence, learning.  
\section{Results}\label{section_results}
\begin{figure*}
	\includegraphics[width=2\columnwidth]{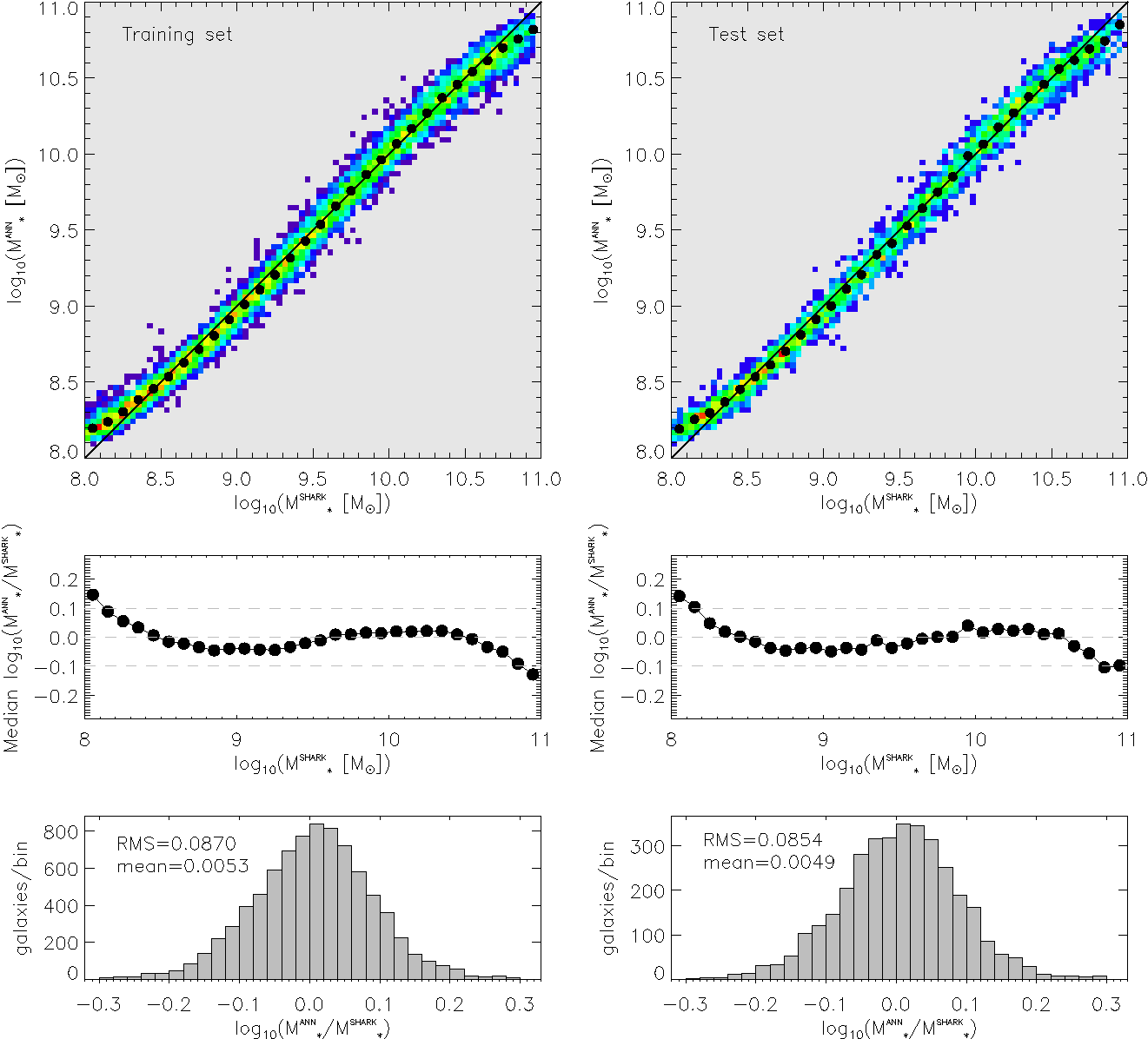}
\caption{Performance summary of the ANN. Left and right columns show results for the training and test sets, respectively. \textbf{Top row:} Two-dimensional distribution of predicted versus \shark-evaluated (i.e., true) stellar masses.  \textcolor{black}{The colour scale represents the number of galaxies per 2D bin, with each bin having a width of 0.0125 dex in both dimensions. The rainbow colour scale progresses from blue (low counts), through green, yellow, and orange, to red (high counts).} The solid black line denotes perfect agreement, while black-filled circles indicate the median predicted stellar mass in 0.1 dex-wide bins of $\log_{10}M_*^\mathrm{SHARK}$. \textbf{Middle row:} Median residuals, $\log_{10}M_*^\mathrm{ANN} - \log_{10}M_*^\mathrm{SHARK}$, as a function of $\log_{10}M_*^\mathrm{SHARK}$, demonstrating consistent accuracy across nearly three decades of stellar mass. \textbf{Bottom row:} Overall residual distributions. These results confirm that the network can reliably estimate stellar masses using only broadband magnitudes and colour indices.}

    \label{results}
\end{figure*}

For the set of \shark-evaluated (i.e., true) stellar masses, \( M_*^\mathrm{SHARK} \), from the training and test data sets, and the corresponding network-predicted stellar masses, \( M_*^\mathrm{ANN} \), the top panels of Figure~\ref{results} display the two-dimensional distribution of \( \log_{10}M_*^\mathrm{ANN} \) versus \( \log_{10}M_*^\mathrm{SHARK} \). The solid black line represents the line of equality, while the black-filled circles indicate the median \( \log_{10}M_*^\mathrm{ANN} \) in bins of width 0.1 dex in \( \log_{10}M_*^\mathrm{SHARK} \). The middle panels show the median offset between predicted and true stellar masses as a function of \( M_*^\mathrm{SHARK} \), and the bottom panels present the overall distribution of residuals, defined as \( \log_{10}M_*^\mathrm{ANN} - \log_{10}M_*^\mathrm{SHARK} \).

The results presented in Fig.~\ref{results} demonstrate that the ANN consistently delivers high-accuracy stellar mass estimates for \shark\ galaxies across the range $8 \lesssim \log_{10}(M_*/M_{\odot}) \lesssim 11$. The network achieves an overall root-mean-square (RMS) error of just $\sim 0.085$~dex in $\log_{10}M_*$ for the test set. However, its performance declines slightly toward the boundaries of the mass range, where the median residuals begin to deviate from zero---specifically for galaxies with $M_* \lesssim 10^{8.4}$~\msun\ or $M_* \gtrsim 10^{10.7}$~\msun. Given the approximately uniform distribution of training-set galaxies in $\log_{10}(M_*^\mathrm{SHARK})$, these reduced accuracies cannot be entirely attributed to a relative lack of training data at the extremes. Rather, they likely reflect an intrinsic limitation of the network in modelling the stellar masses of galaxies at the low- and high-mass ends, as opposed to its robust performance for the majority of galaxies within the intermediate mass range.  \textcolor{black}{For example, regression models trained with quadratic loss functions tend to exhibit regression toward the mean, systematically underpredicting high values and overpredicting low ones.}  Nevertheless, even the largest discrepancies remain well within the uncertainty margins typically associated with other stellar mass estimation techniques, highlighting the ANN's strong overall reliability.

\section{Discussion}\label{section_discussion}
Numerous previous studies have employed neural networks to predict galaxy properties, including stellar mass. For instance, \citet{chu_2024} utilised a multi-branch convolutional neural network applied to galaxies from the TNG100 simulation, combining synthetic $r$-band images, velocity maps, and various galaxy parameters to predict central stellar masses with RMS errors as low as 0.04~dex. Likewise, \citet{zeraatgari_2024} trained a deep learning model on SDSS and WISE magnitudes from the MPA--JHU DR8 catalogue for galaxies with $z<0.3$, achieving an RMS error of 0.206~dex. The present study is distinctive in that it trains a comparatively simple network to produce highly accurate stellar mass estimates using only magnitudes and colours--- directly observable quantities ---without relying on additional structural or kinematic information.

A natural extension of this work is to apply the ANN to predict stellar masses of real galaxies, such as those from the GAMA\footnote{Galaxy and Mass Assembly \citep{GAMA}.} survey, which provides broad-band photometric data ranging from the far-ultraviolet to the mid- and far-infrared. A successful application of the ANN in this context would effectively bridge state-of-the-art theoretical models of galaxy formation and evolution (i.e., from \shark) with high-quality observational data, enabling accurate and efficient stellar mass estimates for large galaxy samples.

\textcolor{black}{In such observational applications, however, the ANN will be confronted with realistic sources of uncertainty—most notably, photometric measurement errors and potential inaccuracies in redshift estimates. These uncertainties can affect the derived colours and absolute magnitudes, thereby introducing scatter and bias in the predicted stellar masses. While this study is based on idealised \shark\ data with perfect photometry and redshift information, future work will need to assess the model’s robustness under observational noise. This could be achieved by retraining and validating the ANN using simulated photometry that has been perturbed according to realistic noise models, and by incorporating photometric redshift errors either directly into the inputs or through marginalisation over redshift probability distributions.}

Traditional methods, such as SED modelling, typically yield stellar mass uncertainties of 0.2–0.3 dex due to degeneracies among stellar age, metallicity, dust attenuation, and star formation history \citep[e.g.,][]{walcher_2011, conroy_2013}. For example, \citet{elson_2024} demonstrates that ALFALFA\footnote{Arecibo Legacy Fast ALFA survey \citep{haynes_2018}.} galaxies with multiple derived stellar mass estimates---whether based on SDSS optical photometry, infrared unWISE photometry, ultraviolet GALEX imaging, or multi-wavelength SED fitting---often exhibit median discrepancies significantly offset from zero, with RMS values spanning several tenths of a dex. Empirical scaling relations offer an alternative approach for estimating stellar masses; however, they typically do not achieve accuracies better than $\sim 0.1$~dex and require precise measurements of various galaxy properties. For instance, for late-type galaxies, the baryonic Tully--Fisher relation yields uncertainties of $\sim 0.1$--$0.4$~dex \citep{bradford_2016, lelli_2016}, while the mass--spin relation can achieve uncertainties as low as $\sim 0.09$~dex \citep{elson_2024}.

\section{Network Dynamics}\label{section_dynamics}
In addition to accurately predicting the evaluated stellar masses of the \shark\ galaxies, the trained ANN also provides insight into the relative importance of the input features. 

Several individual features on their own are found to be reasonably well-correlated with stellar mass. For example, individual magnitudes from the SDSS, VISTA, and WISE bands exhibit linear correlations with stellar mass at the $\sim5$~per~cent level, with RMS uncertainties ranging from $\sim0.4$~dex down to $\sim0.13$~dex (in the case of the VISTA $K_s$ band). Combining these magnitude features enables the network to achieve $\log_{10}M_*$  predictions with a slightly better RMS error of $\sim0.12$~dex. The inclusion of colour indices, however, allows the full network to more precisely distinguish between different galaxy masses, leading to the enhanced predictive performance observed (i.e., $\mathrm{RMS}\approx0.085$~dex for the test set).

An approach to understanding the network's overall dynamics is to quantify how the predicted stellar mass changes with respect to each input feature for different types of galaxies. This can be done using an analytical expression. Let $f = \log_{10}M_*^\mathrm{ANN}$ denote the predicted logarithmic stellar mass from the ANN for a galaxy with a set of input features $\vec{x} = [x_0, x_1, x_2, \dots, x_{28}]$. The partial rate of change of $f$ with respect to feature $x_i$ is given by
\begin{equation}\label{BFE}
\frac{\partial f}{\partial x_i} = \sum_{j=0}^{m} \frac{w_j^\mathrm{out} \times w_{ji}^\mathrm{HL} \times e^A}{(1 + e^A)^2},
\end{equation}
where $m = 15$ is the number of hidden-layer neurons, and
\begin{equation}
A = -\sum_{i=0}^{n} w_{ji}^\mathrm{HL} \times x_i - b_j^\mathrm{HL} ,
\end{equation}
with $n = 28$ the number of input features\footnote{A derivation of Eqn~\ref{BFE} can be found in Appendix~\ref{AppA}.}. Thus, for any feature of any galaxy, the partial derivative $\frac{\partial f}{\partial x_i}$ can be evaluated given the trained network's weight and bias parameters.

\begin{figure*}
	\includegraphics[width=2\columnwidth]{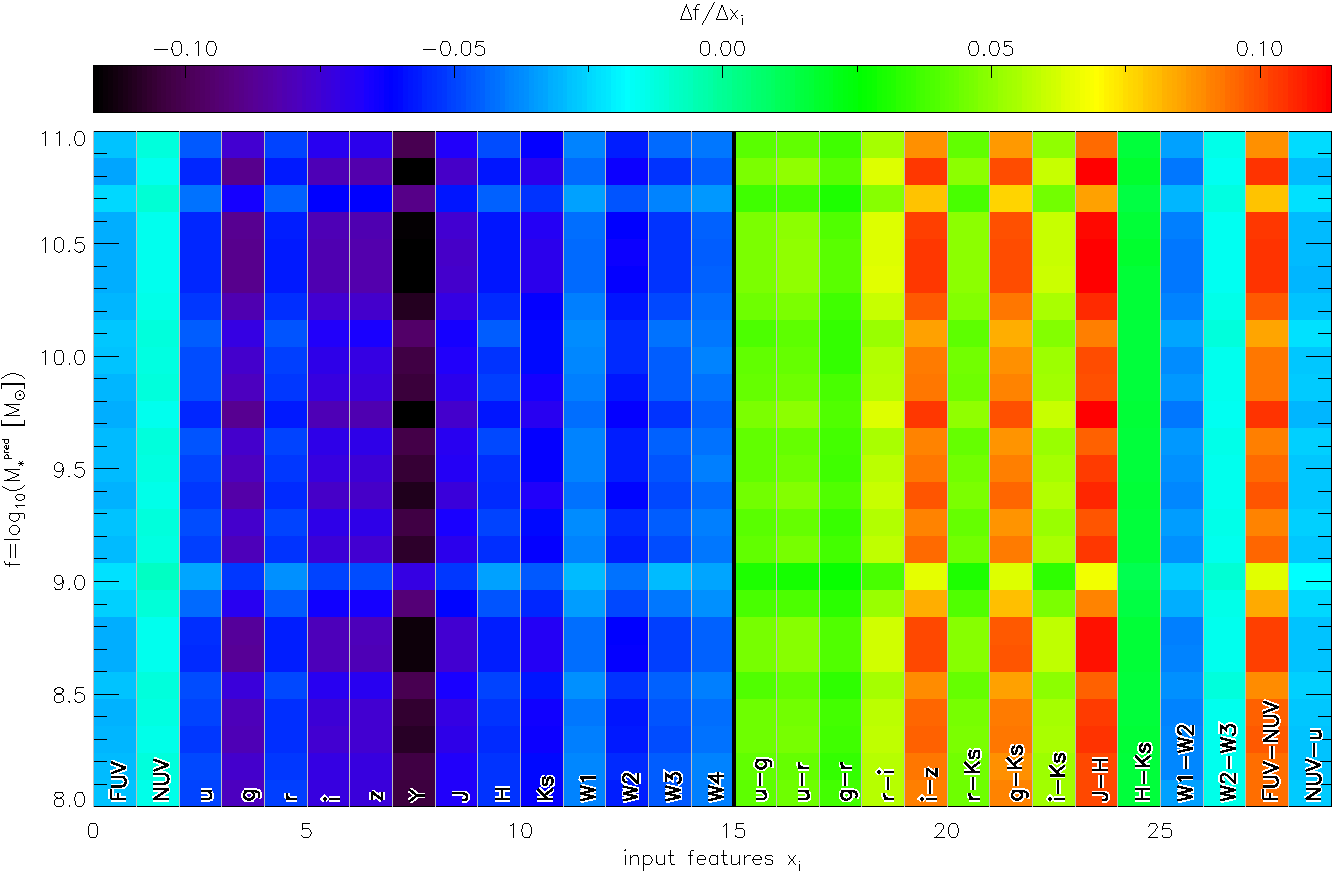}
	\caption{Distribution of the partial derivatives ${\partial f}/{\partial x_i}$, where $f = \log_{10}(M_*^\mathrm{ANN}~[M_{\odot}])$, shown as a function of both the predicted stellar mass $f$ and the corresponding input feature $x_i$.  Each row corresponds to a spectrum in Fig.~\ref{spectra}.  The vertical black line separates the input features into magnitudes (to the left) and colour indices (to the right)}
	\label{dfdx}
\end{figure*}

For the set of galaxies whose broad-band spectra are shown in Fig.~\ref{spectra}, the distribution of ${\partial f}/{\partial x_i}$ values in the $f$–$x_i$ plane is presented in Fig.~\ref{dfdx}.  Each row in the figure corresponds to one of the galaxies. The vertical black line separates the input features into magnitudes (to the left) and colour indices (to the right).  In general, magnitudes and colours contribute to stellar mass predictions in opposing ways: magnitudes tend to reduce the predicted mass, while most colour indices increase it. Among the most influential colours are $i-z$, $g-K_s$, and $J-H$, each capable of increasing a galaxy’s predicted $f$ by up to $\sim 0.1$~dex. These indices are sensitive to variations in the galaxy’s SED driven by differences in stellar age, metallicity, and the mix of low- and high-mass stars, as well as the integrated light from evolved stellar populations. The $FUV-NUV$ colour, which traces recent or ongoing star formation, also emerges as a key predictive feature. Notably, the network appears to have implicitly learned to accommodate dust attenuation effects in the relevant bands, further enhancing the usefulness of this index. Among the magnitudes, VISTA-$Y$ stands out as the most significant negative contributor to $f,$ exerting the strongest downward influence on predicted stellar mass. Most of the remaining magnitudes exert a similar suppressive effect, with SDSS magnitudes having a slightly greater impact overall.

\begin{figure*}
	\includegraphics[width=2\columnwidth]{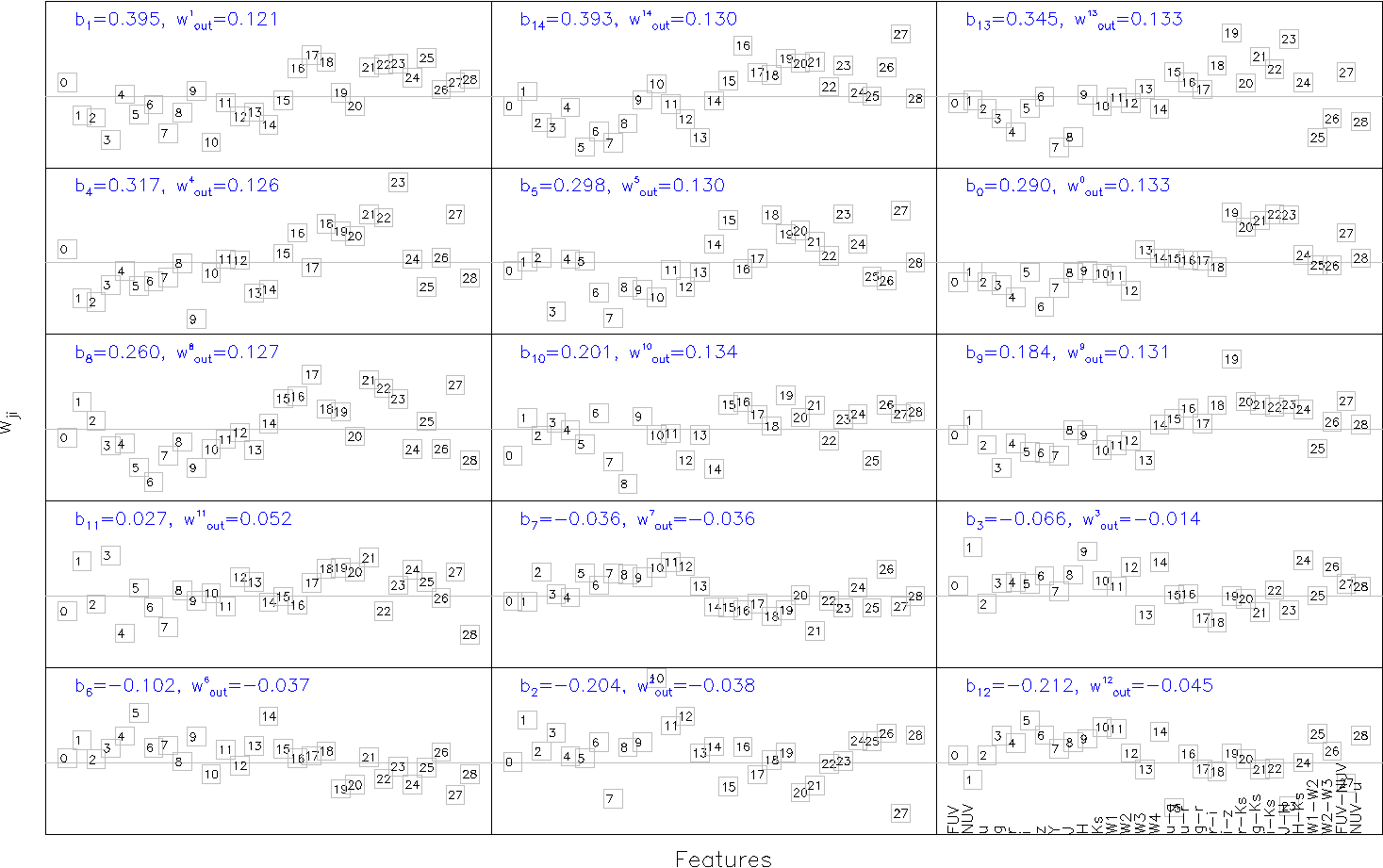}
	\caption{Visualisation of the ANN’s bias and weight parameters. Each panel corresponds to one hidden-layer neuron, ordered by decreasing bias magnitude (e.g., neuron 1, with a bias of 0.395, is shown in the first panel). The weight connecting each neuron to the ReLU output is also indicated. Axes are consistent across all panels, with $y$-axes spanning the range $-0.59$ to $0.79$. The horizontal grey line in each panel denotes zero. For each hidden-layer neuron, the 29 associated weights correspond to input features are labelled by number: 0–14 represent absolute magnitudes in GALEX ($FUV$, $NUV$), SDSS ($u$, $g$, $r$, $i$, $z$), VISTA ($Y$, $J$, $H$, $Ks$), and Spitzer ($W1$–$W4$) bands; 15–28 denote colour indices $u-g$, $u-r$, $g-r$, $r-i$, $i-z$, $r-Ks$, $g-Ks$, $i-Ks$, $J-H$, $H-Ks$, $W1-W2$, $W2-W3$, $FUV-NUV$, and $NUV-u$. The final panel provides a key to the feature labels. The visualisation highlights that colour indices are among the most influential features in the network's estimation of stellar mass.}
	\label{flow}
\end{figure*}

The network can also be analysed on a parameter-by-parameter basis.  Figure~\ref{flow} visualises the ANN’s parameters on a neuron-by-neuron basis. Each panel corresponds to a single neuron from the hidden layer, arranged from left to right, top to bottom, in order of decreasing bias parameter— an important value that determines how easily and strongly the neuron is activated. For example, hidden-layer neurons 1, 14, and 13 are the most active, with bias values of 0.395, 0.393, and 0.345, respectively. In contrast, neurons 6, 2, and 12 (shown in the bottom row) are the least active. Each panel also displays the weight linking the hidden-layer neuron to the ReLU activation neuron in the output layer. A hidden-layer neuron with both a high bias parameter and a strong connection to the output layer represents a key pathway through which information can flow from the input features to the final prediction.  

The amount of information each hidden-layer neuron receives from the input layer is governed by the relative importance assigned to each feature. Accordingly, each panel in Fig.~\ref{flow} displays the weights connecting a given hidden-layer neuron to the 29 input features, offering a visual summary of which features most strongly influence that neuron's activation. When these weight patterns are considered alongside each neuron's bias and output weight, a consistent trend emerges: neurons that contribute most effectively to the final ReLU output tend to assign greater weight to specific colour indices. For example, hidden-layer neuron~1 places particular emphasis on $g-r$, $r-i$, $g-K_s$, $i-K_s$, $J-H$, and $W1-W2$.  When examining hidden-layer neurons with lower bias parameter values---which require stronger weighted inputs to activate significantly (e.g., those shown in the last two rows of Fig.~\ref{flow})---it becomes evident that both magnitudes and colour indices together play a crucial role in determining stellar mass. Magnitudes are used to identify galaxies with atypical spectral shapes, thereby enabling the network to capture a broader diversity of galaxy characteristics.

Ultimately, these complex, non-linear combinations of input features generated by individual neurons and illustrated in Fig.~\ref{flow} allow the network to detect important patterns in the data, leading to accurate predictions of stellar mass.

\section{Conclusions}\label{section_conclusions}
This study has shown that a simple, fully connected artificial neural network (ANN) with a single hidden layer can accurately predict the stellar masses of star-forming galaxies using only broadband magnitudes and colour indices. Trained on simulated galaxies from the \shark\ semi-analytic model, the ANN recovers the true stellar masses of a test set with an RMS error of just 0.085~dex across a stellar mass range spanning $10^8$–$10^{11}$\msun. The median discrepancy remains close to zero over much of this range, increasing to $\gtrsim 0.1$~dex only at the lower and upper extremes. This level of accuracy compares favourably with traditional SED fitting techniques, which often yield uncertainties of 0.2–0.3~dex due to degeneracies in stellar population parameters.

A key outcome of this work is the identification of specific colour indices that play a dominant role in stellar mass estimation. Analyses of both the overall network behaviour and individual neuron dynamics provide quantitative evidence for the predictive power of various indices. While optical and mid-infrared colours are shown to be important, the $FUV-NUV$ index also emerges as a significant contributor, suggesting that the network has implicitly learned to account for attenuation effects in the ultraviolet bands—thereby enhancing the reliability of this feature. The combined use of galaxy magnitudes and colour indices as input features enables the network to detect meaningful patterns in the data, leading to accurate mass predictions across a wide range of galaxy properties.

The simplicity of the network architecture, coupled with its strong predictive performance and low computational demands, highlights the transformative potential of machine learning in astrophysics. In particular, the sort of trained network developed in this study paves the way for a new paradigm in estimating stellar masses—one that shifts away from traditional empirical calibrations or spectral energy distribution fitting and instead relies on directly applying our best theoretical understanding from cosmological simulations. By deploying artificial neural networks trained on state-of-the-art models (e.g., the \shark\ simulations) to the panchromatic broad-band photometry of observed galaxies (such as those from the GAMA survey), this paradigm enables the agnostic and accurate prediction of stellar masses directly from observations. This approach represents a significant step toward unifying theory and observation in the era of data-intensive astronomy.

\section*{Acknowledgements}

\textcolor{black}{The author thanks the anonymous referee for their critical and constructive feedback, which significantly improved the overall quality of this paper.}
\section*{Data Availability}
The parameters of the trained network will be made available upon reasonable request.



\bibliographystyle{rasti}


\appendix
\section{Derivation of Equation~\ref{BFE}}\label{AppA}

\begin{figure*}
	\includegraphics[width=2\columnwidth]{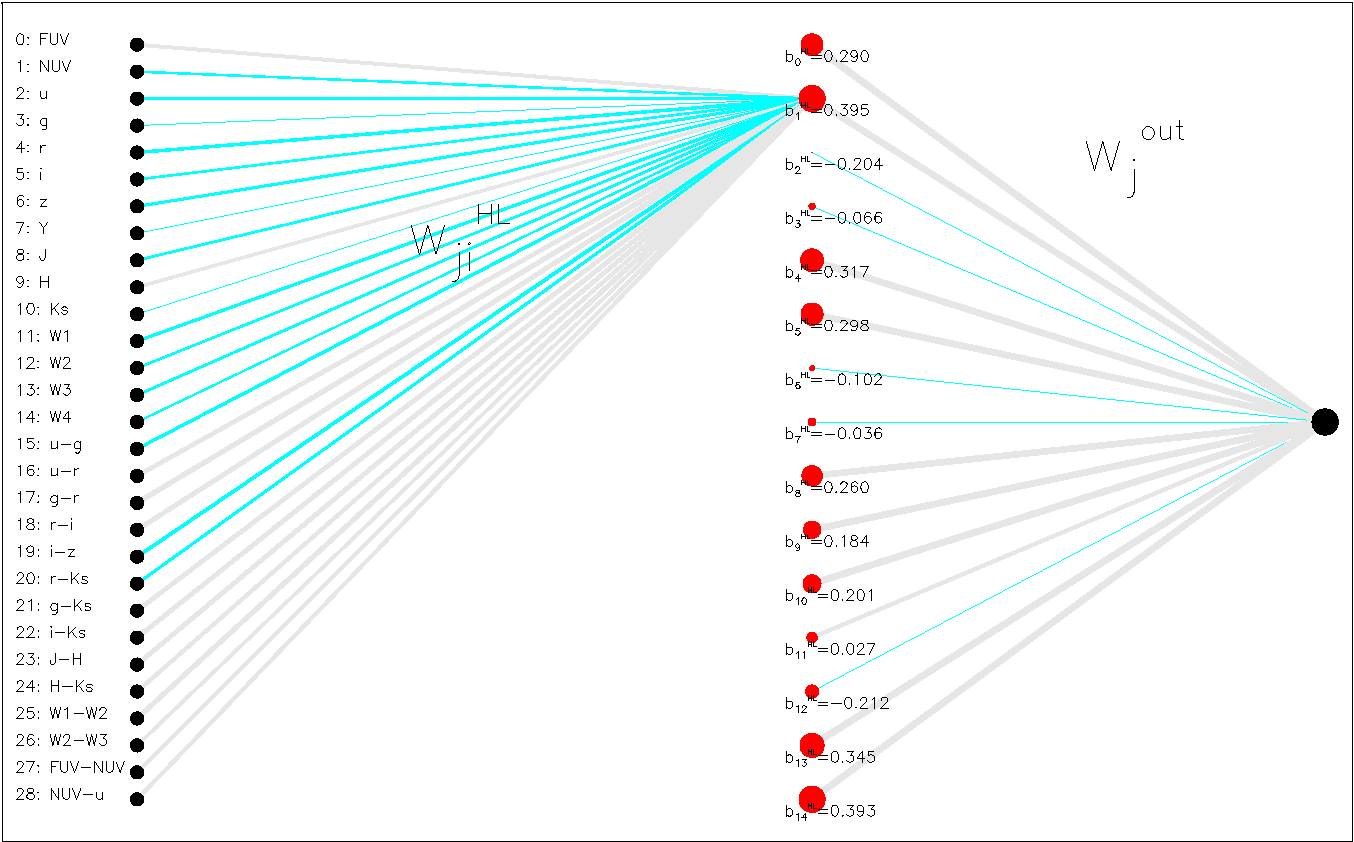}
   \caption{Graphical representation of the ANN developed and trained in this study.  The input layer consists of 29 nodes (shown as black-filled circles), each corresponding to a photometric feature of a galaxy. These features are listed on the left-hand side of the figure: indices 0–14 denote absolute magnitudes in the GALEX ($FUV$, $NUV$), SDSS ($u$, $g$, $r$, $i$, $z$), VISTA ($Y$, $J$, $H$, $Ks$), and Spitzer ($W1$–$W4$) bands, while indices 15–28 represent colour indices: $u-g$, $u-r$, $g-r$, $r-i$, $i-z$, $r-Ks$, $g-Ks$, $i-Ks$, $J-H$, $H-Ks$, $W1-W2$, $W2-W3$, $FUV-NUV$, and $NUV-u$. The network's 15 hidden-layer neurons are shown as red-filled circles, with their sizes proportional to the bias parameter of each neuron (explicitly listed below each one). The synapses $w_{1,i}^\mathrm{HL}$ linking the input nodes to hidden-layer neuron 1—the neuron with the largest bias parameter—are shown in the left portion of the figure. Synapse thickness is linearly proportional to the weight value, with cyan lines indicating negative weights and grey lines indicating positive weights. While all input nodes connect to every hidden-layer neuron, only the connections to neuron 1 are shown here for clarity. Finally, the synapses connecting the hidden-layer neurons to the output ReLU neuron are similarly illustrated.}
    \label{ANN_fig}
\end{figure*}

A derivation of Eqn~\ref{BFE}, which expresses how changes in the input features affect the output of the network’s ReLU neuron, is presented below. Readers may find it helpful to refer to Fig.~\ref{ANN_fig}, which provides a graphical representation of the network architecture and may aid in following the derivation.

Given the network’s 29 input features—each represented by a node in the input layer and indexed by $i = 0,~1,~2,~\ldots,~28$—and its 15 hidden-layer neurons indexed by $j = 0,~1,~2,~\ldots,~14$, the weighted sum of inputs to the $j$-th hidden-layer neuron is
\begin{equation}
\sum_{i=0}^{n} w_{ji}^\mathrm{HL} x_i,
\end{equation}
where $w_{ji}^\mathrm{HL}$ are the connection weights and $x_i$ are the input feature values. Adding the bias term $b_j^\mathrm{HL}$ yields the argument $z_j^\mathrm{HL}$ for the neuron’s activation function:
\begin{equation}
\sigma_j^\mathrm{HL} = \frac{1}{1 + \exp\left(-z_j^\mathrm{HL}\right)} = \frac{1}{1 + \exp\left(-\sum_{i=0}^{n} w_{ji}^\mathrm{HL} x_i - b_j^\mathrm{HL}\right)}.
\end{equation}

Given the connections (with weights \(w_j^\mathrm{out}\)) from the hidden-layer neurons to the network's output ReLU neuron, the activated output can be expressed as
\[
f = b^\mathrm{out} + \sum_{j=0}^{m} w_j^\mathrm{out}\,\sigma_j^\mathrm{HL}.
\]

An expanded expression for \(f\) is
\begin{eqnarray}
f &=& b^\mathrm{out} \nonumber \\
  && +\, \frac{w_0^\mathrm{out}}{1 + \exp\left(-\sum_{i=0}^{n} w_{0i}^\mathrm{HL} x_i - b_0^\mathrm{HL}\right)} \nonumber \\
  && +\, \frac{w_1^\mathrm{out}}{1 + \exp\left(-\sum_{i=0}^{n} w_{1i}^\mathrm{HL} x_i - b_1^\mathrm{HL}\right)} \nonumber \\
  && +\, \cdots \nonumber \\
   && +\, \frac{w_j^\mathrm{out}}{1 + \exp\left(-\sum_{i=0}^{n} w_{ji}^\mathrm{HL} x_i - b_j^\mathrm{HL}\right)}\nonumber \\
     && +\, \cdots \nonumber \\
  && +\, \frac{w_m^\mathrm{out}}{1 + \exp\left(-\sum_{i=0}^{n} w_{mi}^\mathrm{HL} x_i - b_m^\mathrm{HL}\right)}.
\end{eqnarray}

To generate an expression for the sensitivity of the network output with respect to the input feature \(x_i\), let's focus on the \(j\)-th term of the sum in the expression for \(f\). This requires evaluating the partial derivative
\begin{equation}
{\partial f\over \partial x_i} = \frac{\partial}{\partial x_i} \left( \frac{w_j^\mathrm{out}}{1 + \exp\left( -\sum_{k=0}^{n} w_{jk}^\mathrm{HL} x_k - b_j^\mathrm{HL} \right)} \right).
\label{jth_term}
\end{equation}

This derivative is obtained using the quotient rule, which states that for differentiable functions \(g(x)\) and \(h(x)\),
\begin{equation}
\left( \frac{g}{h} \right)' = \frac{g'h - gh'}{h^2}.
\end{equation}
Applying this to the expression in Eqn~\ref{jth_term}, where the numerator and denominator are identified as \(g = w_j^\mathrm{out}\) (a constant) and \(h = 1 + \exp\left(-\sum_{k=0}^{n} w_{jk}^\mathrm{HL} x_k - b_j^\mathrm{HL}\right)\), respectively, let's define
\[
A = -\sum_{k=0}^{n} w_{jk}^\mathrm{HL} x_k - b_j^\mathrm{HL}.
\]
Then, $h'=-w_{ji}\times e^A$ and the partial derivative of the network output with respect to the input feature \(x_i\) can be expressed as
\begin{equation}
\frac{\partial f}{\partial x_i} = \sum_{j=0}^{m} \frac{w_j^\mathrm{out} \, w_{ji}^\mathrm{HL} \, e^{A}}{(1 + e^{A})^2} ,
\end{equation}
This expression highlights how each feature influences the output through the product of hidden-layer and output-layer weights.


\bsp	
\label{lastpage}
\end{document}